\documentclass[12pt, preprint]{aastex}

\newcommand{\avg}[1]{\left\langle#1\right\rangle}
\newcommand{\rp}[1]{\left(#1\right)}
\newcommand{\Om}{\Omega_m}
\newcommand{\lbox}{\mathrm{L_{box}}}

\newcommand{\Lam}{\Lambda}

\newcommand{\hinv}{{h^{-1}}}
\newcommand{\mpc}{{\rm\,Mpc}}

\newcommand{\himpc}{\hinv{\rm\,Mpc}}

\newcommand{\Lsunb}{L_{\odot,B}}

\newcommand{\etal}{et~al.}

\newcommand{\Fig}[1]{Figure~\ref{#1}}
\newcommand{\beq}{\begin{equation}}
\newcommand{\eeq}{\end{equation}}
\newcommand{\drho}{\delta\rho}

\slugcomment{Accepted to ApJ}

\shorttitle{PDF of Light}
\shortauthors{Ostriker et al.}

\begin{document}

\title{The Probability Distribution Function of Light in the Universe: 
Results from Hydrodynamic Simulations}

\author{Jeremiah P. Ostriker} 
\affil{Institute of Astronomy, University of Cambridge, Madingley Road, Cambridge, CB3, OHA, UK}
\email{jpo@ast.cam.ac.uk}

\author{Kentaro Nagamine}
\affil{Harvard-Smithsonian Center for Astrophysics, 60 Garden Street,  
Cambridge, MA 02138, U.S.A.}
\email{knagamin@cfa.harvard.edu}

\author{Renyue Cen} 
\affil{Princeton University Observatory, Princeton, NJ 08544, U.S.A.}
\email{cen@astro.princeton.edu}

\and

\author{Masataka Fukugita}
\affil{Institute for Cosmic Ray Research, University of Tokyo, 
Kashiwa 2778582, Japan}
\email{fukugita@icrr.u-tokyo.ac.jp}

%%%%%%%%%%%%%%%%%%%%%%%%%%%%%%%%%%%%%%%%%%%%%%%%%%%%%%%%%%%%%%%%%%%%%%%%

\begin{abstract} 
While second and higher order correlations of the light distribution 
have received extensive study, the lowest order probability distribution 
function (PDF) --- the probability that a unit volume of space will 
emit a given amount of light --- has received very little attention. 
We estimate this function with the aid of hydrodynamic simulations of the 
$\Lam$CDM model, finding it significantly different from the mass density PDF, 
and not simply related to it by linear bias or any of the other
prescriptions commonly adopted.  If the optical light PDF is, in reality, 
similar to what we find in the simulations, then some measures of $\Om$ based 
on mass-to-light ratio and the cosmic virial theorem will have significantly 
underestimated $\Om$. Basically, the problem is one of selection bias, with 
galaxy forming regions being unrepresentative of the dark matter distribution
in a way not described by linear bias. Knowledge of the optical PDF 
and the plausible assumption of a log-normal distribution 
for the matter PDF will allow one to correct for these selection 
biases. We find that this correction (which amounts to $20-30\%$) 
brings the values of $\Om$ estimated by using the mass-to-light ratio 
and the cosmic virial theorem to the range $\Om=0.2-0.3$, 
in better agreement with the WMAP result than the uncorrected estimates.
In addition, the relation between mass and light PDFs gives us insight
concerning the nature of the void phenomenon. In particular our 
simulation indicates that 20\% of mass is distributed in voids,
which occupy 85\% of volume in the universe.
\end{abstract}


%%%%%%%%%%%%%%%%%%%%%%%%%%%%%%%%%%%%%%%%%%%%%%%%%%%%%%%%%%%%%%%%%%%%%%

\section{Introduction} 
\label{Section:Intro}

 In modern cosmological studies, the distribution of the
matter density has received extensive study, based on theory and on
increasingly accurate dark matter numerical simulations. 
We may study the matter distributions directly or by looking at the
distribution of halos after applying group finding
algorithms to identify virialized ``halos'' in $N$ body simulations. 
The simplest statistic is 
the probability of a sphere of radius $R$ to contain
mass in the range $M~-~ (M+dM)$ or, alternatively, 
to have a density in the range $\rho_m~-~ (\rho_m + d\rho_m)$,
dividing by the volume of the sphere.
We may more conveniently use the normalized dimensionless density, 
$y \equiv \rho_m / \avg{\rho_m}$, as the parameter, 
further dividing by the mean mass density, and
consider the probability distribution function (PDF) of $y$. 
By the definition of a PDF, 
\beq
\int_0^\infty f_R(y)dy = 1,
\eeq
and the commonly used density variance is
\beq
\sigma_R^2\equiv \int_0^\infty (y-1)^2f_R(y)dy \equiv \avg{(y-1)^2}_R 
\equiv \avg{\delta_m^2}_R,
\eeq
where the additional definitions are given to establish nomenclature
used in this paper, and the subscript $R$ shows the dependence of 
the averaged quantities on scale $R$ (we usually drop $R$, however, 
unless we need to emphasize it).

It is known in CDM models that the logarithm of the density 
shows a normal distribution to a good approximation when the structure
formation has reached the non-linear regime \citep{Coles91, Kofman94, 
Taylor00, Kayo01}:
\beq
\label{Eq:lognormal}
f(y) = \frac{1}{\sqrt{2\pi\omega^2}} \frac{1}{y} \exp\left(
-\frac{[\ln(y)+\omega^2/2]^2}{2\omega^2} \right).
\eeq
\noindent
The parameter that specifies the distribution, $\omega$, is 
related to the variance $\sigma$ of the density field as,
\beq
\label{Eq:omega}
\omega_R^2 = \ln(1+\sigma_R^2).
\eeq
It is known that this log-normal function can be obtained from 
the one-to-one mapping between the linear random Gaussian field and the
nonlinear density field \citep{Coles91}, although the physical
meaning of the transformation is not well understood. 
The mapping, nevertheless, seems to capture an important 
aspect of the nonlinear evolution of the density field in the universe.

In parallel to the PDF of the mass distribution $f(y)$,
we also define the PDF of the light distribution by $g(j)$, 
where $j$ is the normalized luminosity density.
It is often convenient to define the logarithm of each 
parameter as $Y\equiv\log y$ and $X\equiv\log j$.

It is somewhat curious that much less attention has been paid 
to the PDF of the observed light in the universe, whereas
two particle and higher order distribution functions have been 
studied extensively in the literature.
The work on ``counts-in-cells'' by \citet{Efstathiou90} 
implicitly deals with the projected mass PDF in the linear regime, and the 
recent work of \citet{Dekel99} studied this function in the context of 
``stochastic biasing''.  The void distribution function, which depends on 
all higher order distributions, has received some attention 
\citep[e.g.][]{Vogeley91, Vogeley94, Benson03}.
However, we are not aware of any systematic analysis 
of the optical PDF or attempts to relate it with the mass PDF
(but see the work of \citet{Sigad00} where they attempted to relate 
the galaxy biasing function and the cumulative distribution functions).

In this paper, we consider the relation between $g(j)$ and $f(y)$,
or more explicitly, the relation between $X$ and $Y$, $X=X(Y)$. 
These relations can be determined by the physics of galaxy
formation, and we illustrate them with the use of hydrodynamic simulations
of the concordance $\Lam$CDM model. The results allow us to elucidate
the nature of the void phenomenon (Kirshner et al. 1981; Rood 1988 and references
therein; Peebles 2001) in the CDM model that is confronted
with the observation. Our results indicate that the relation between $X$
and $Y$ is not linear, contrary to what has usually been assumed.
This also points to the presence of systematic biases, when galaxies 
are used as tracers of the mass, in the conventional estimates for 
the total mass density of the universe and for the evaluation of 
the cosmic virial theorem. Our results may be used to correct for 
these biases. 
 
We show in Section \ref{Section:Result} the PDF of the mass and light
distributions and their relation to each other using a hydrodynamic 
simulation. In Section \ref{Section:Applications} we discuss implications 
to observations, in particular, how selection bias can be corrected
with the knowledge of these functions.

%%%%%%%%%%%%%%%%%%%%%%%%%%%%%%%%%%%%%%%%%%%%%%%%%%%%%%%%%%%%%%%%%%%%%%

\section{PDF from Hydrodynamic Simulations}
\label{Section:Result}

\subsection{Simulation}

The simulation we use is similar to \citet{CO93}, but has been 
significantly improved over the years by adding more elaborate 
modeling. The simulation which uses the Total Variation Diminishing (TVD) 
method \citep{Ryu} is performed for a box-size of $\lbox=25\himpc$ 
with Eulerian hydrodynamic mesh of $768^3$ and cosmological parameters of 
$(\Om, \Omega_{\Lam}, \Omega_b, h, \sigma_8)=(0.3, 0.7, 0.035, 0.67, 
0.9)$, which are close to the estimate by WMAP \citep{Spergel03}. 
This is the same simulation that has been extensively 
used in our earlier papers \citep{Nagamine01a, Nagamine01b, 
Nagamine02} to discuss star formation history, luminosity function, 
and the nature of Lyman-break galaxies.
We refer to \citet{CO02} and \citet*{Nagamine01b} for more details 
of the simulation.
As in our previous analyses we employ the population synthesis model of 
GISSEL99 (Bruzual \& Charlot 1993; Charlot 1999, private communication) 
to calculate the stellar luminosity. Unless otherwise noted, all the 
luminosity used in this paper is in the $B$ band, neglecting the dust 
extinction effect.

%%%%%%%%%%%%%%%%%%%%%%%%%%%%%%%%%%%%%%%%%%%%%%%%%%%%%%

\subsection{One-Point PDF of Mass and Light}

We calculate mass and light contained in spheres 
of radius $R$ located at arbitrary locations, i.e., smoothed with
a tophat filter.  We then compute the probability $dP$ of a 
sphere to have the normalized density parameter $y~-~ (y+dy)$ for mass 
and $j~-~ (j+dj)$  for light: 
$dP = f(y)dy = g(j)dj$, where $f(y)$ and $g(j)$ are the one-point
PDF of mass and light, respectively.

In \Fig{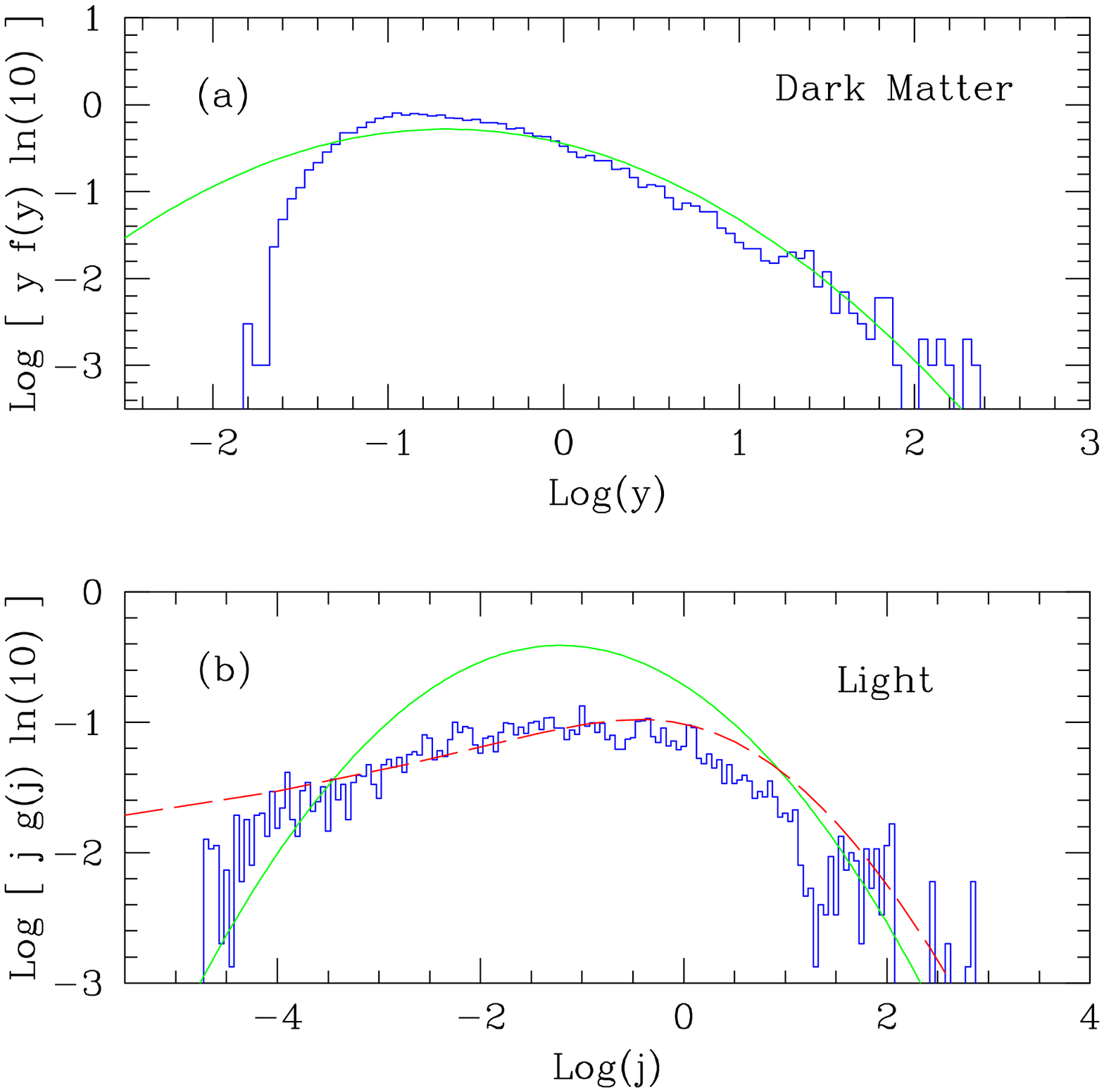}, we show the one-point PDFs $f(y)$ in panel (a) 
and $g(j)$ in panel (b) computed from our simulation  
with $R=1\himpc$, a scale which is small enough to be 
sensitive to the local galaxy formation environment but large enough 
for the simulation to be accurately representing physical values. 
Since our binning is done in $\log y$, the plotted histogram actually 
corresponds to $(\ln10)~yf(y)$ [$f(y)dy=(\ln10)~yf(y)d\log y$], and 
similarly for $g(j)$.  Note the difference in the scales of 
the abscissas of the two panels.

The solid curves plotted over the histogram in both panels are 
log-normal distribution with the variance parameter $\omega^2$ computed 
from the same simulation, $\omega^2= 3.1$ for dark matter 
and $\omega^2= 5.6$ for light.
The agreement between the computed dark matter PDF and the 
log-normal distribution is reasonably good.
The departure in our computation, especially at small $\log y$, 
is ascribed to the limited box-size of our simulation.  
Note that the light PDF strongly deviates from the log-normal distribution.
(The {\it long-dashed} line in \Fig{f1.eps}(b) 
will be explained in Section~\ref{Section:XYrelation}.)

%%%%%%%%%%%%%%%%%%%%%%%%%%%%%%%%%%%%%%%%%%%%%%%%%%%%%%%%%%%%

\subsection{Relation between $\log y$ and $\log j$}
\label{Section:XYrelation}

\Fig{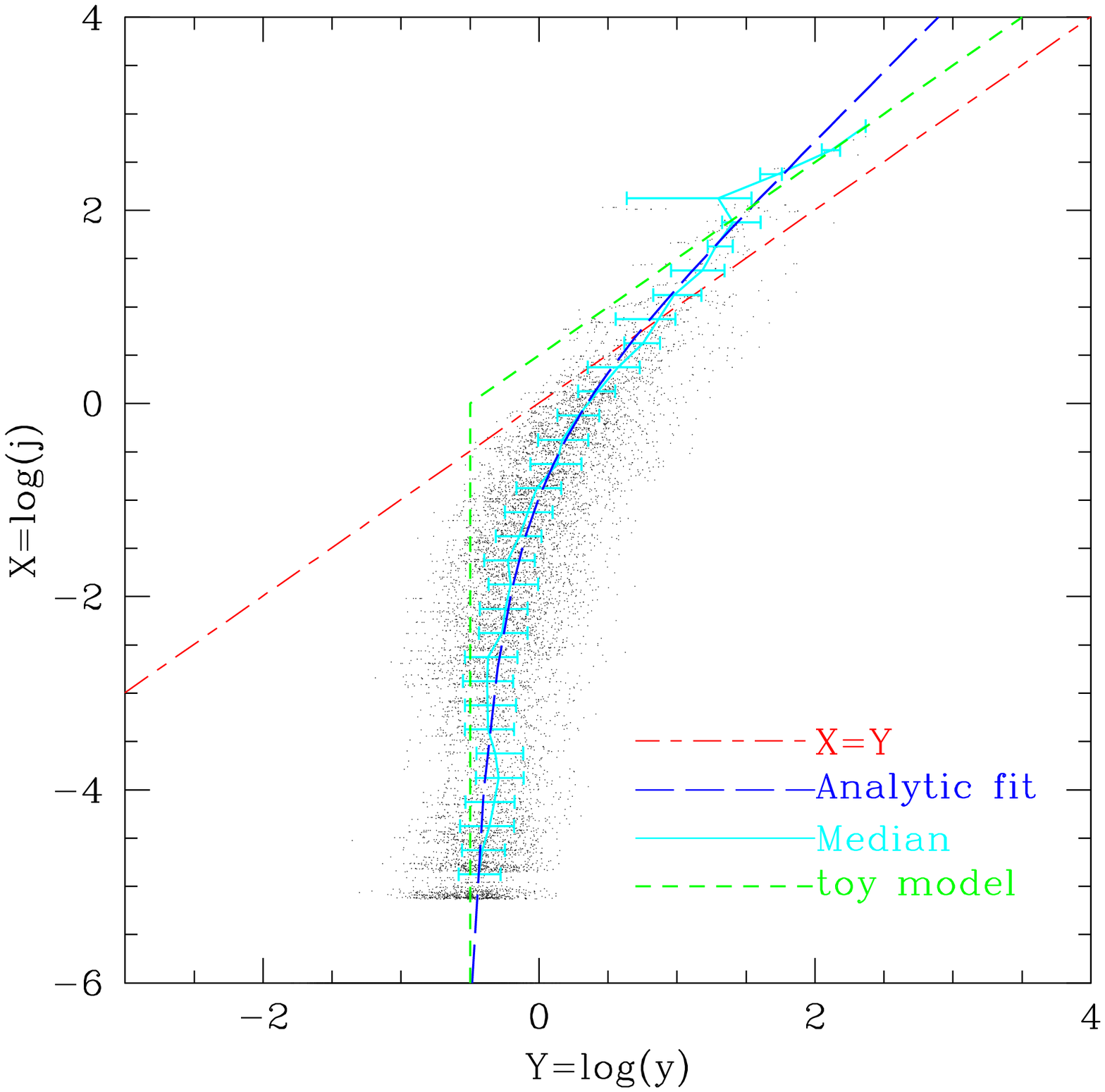} shows the light density parameter 
$X\equiv\log j$ against the dark matter density parameter 
$Y\equiv\log y$. \footnote{We note that a similar figure was reported 
by \citet{Blanton99} from our earlier simulations, but they used
stellar mass density instead of $j$. The figure was also  
truncated at $(\log y, \log j)\approx (0.5, 0.5)$. The simulation 
box-size we use here is smaller than that Blanton et al. used, and
does not contain massive clusters that show the suppression 
of luminosity density at the highest mass density region.}  
The point in the figure represents mass and light contained in 
each sphere. 
The {\it solid} line shows the median value in each $X$ bin, with 
error bars in $X$-direction being the quartiles on both sides.  
The {\it long-short dashed} line shows the identical regression 
$X=Y$, and the {\it long-dashed} line shows an empirical fit
to the simulation data:
\beq
\label{Eq:fit}
X = a Y - \frac{b}{(Y-c)^2}, \quad (Y>c).
\eeq
The parameter $c$ corresponds to an effective cutoff of the mass 
density, below which the sphere does not contain any galaxies that emit
light. The figure indicates that $c\approx -0.5$, i.e., the region of
mass density smaller than 1/3 of the mean is dark. 
This dark region occupies approximately 60\% of the volume when 
smoothed with a tophat sphere of $R=1\himpc$, and contains a fraction 
$\sim 8\%$ of the total mass (these fractions are larger for voids that
are operationally defined in observations: these numbers will be derived
in Section~\ref{Section:distribution}). 
This is identified as the  ``void phenomenon'' in the CDM universe.
Luminous galaxies reside only in the dense regions. Sub-luminous
galaxies prefer to lie in relatively low density regions, but
the sharp increase of $\log j$ at around $Y\approx c$ means that 
they live only in the edge of the voids, certainly not in the
middle of the voids.  These trends seem to be qualitatively consistent
with the observed void phenomenon \citep{Peebles01}.

\begin{figure}
\epsscale{1.0}
\plotone{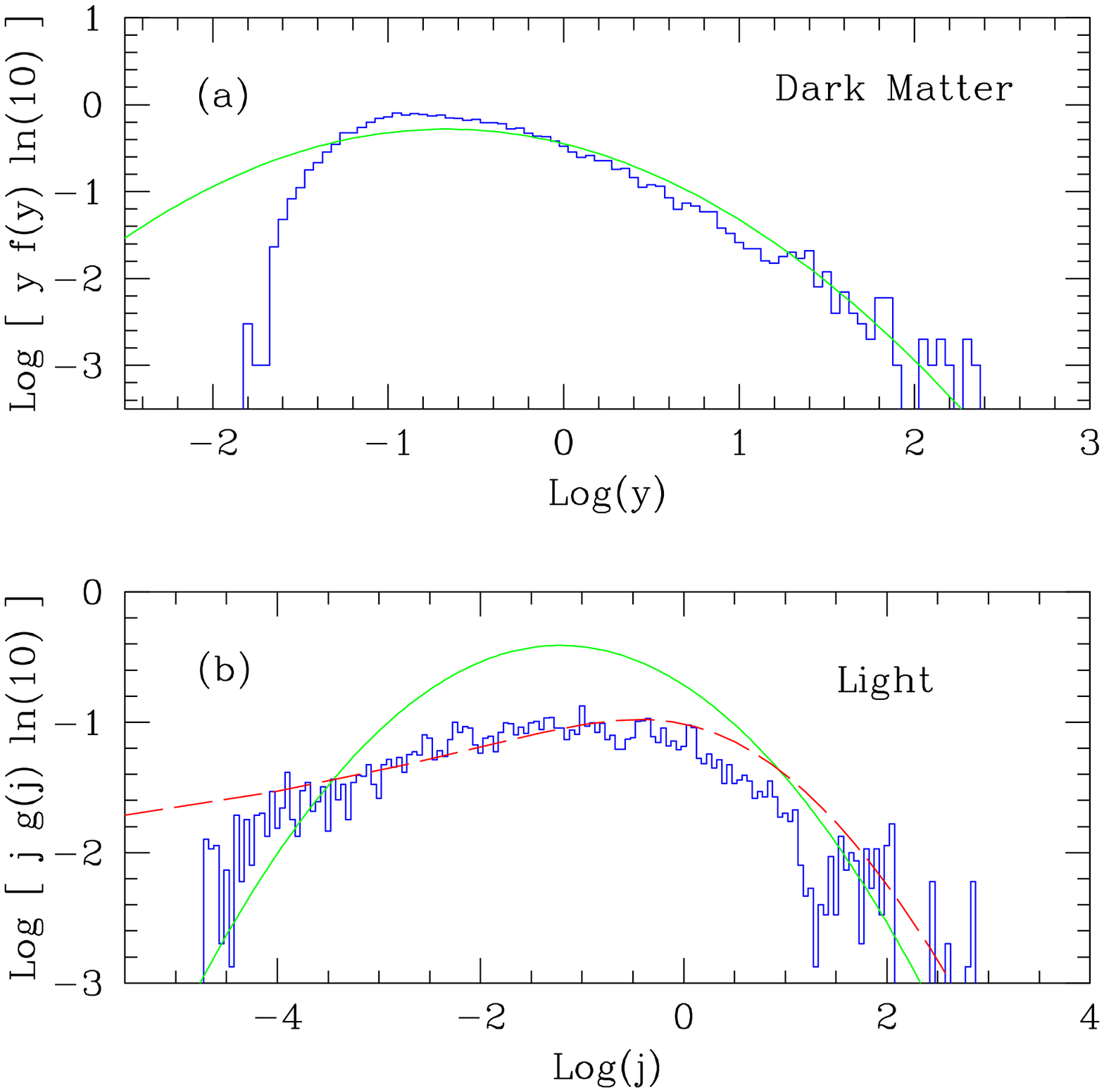}
\caption{One-point PDFs of dark matter $f(y)$ (panel (a)) and of 
light $g(j)$ (panel (b)) computed from a cosmological simulation 
of box-size $25\himpc$ with a tophat window of size $R=1\himpc$.
Note that the plotted histogram actually corresponds to $(\ln10) yf(y)$ 
because the binning is done in $\log(y)$ instead of linear $y$.
The {\it solid} curve in both panels is the log-normal fit with the 
variance parameter $\omega^2=3.1$ (for dark matter) and $5.6$ (for light)
[see eqs. (\ref{Eq:lognormal}) and (\ref{Eq:omega})] computed from the 
simulation.  The {\it long-dashed} line in the lower panel is computed 
by assuming the log-normal distribution for $f(y)$ and the analytic fit 
to the relation between $\log y$ and $\log j$ as shown in \Fig{f2.eps} 
and described in \S~\ref{Section:XYrelation}.  
\label{f1.eps}}
\end{figure}

Given the relation between $X$ and $Y$ in Equation~(\ref{Eq:fit}), 
and assuming a log-normal distribution for $f(y)$ which is specified 
by $\omega$, one can compute the PDF of light via $g(j) \equiv f(y)dy/dj$. 
The {\it long-dashed} curve given in \Fig{f2.eps} is a fit by 
$a=1.4$, $b=0.7$, and $c=-0.85$ for the range $-0.85\leq Y \leq 2.5$
(the actual value of $c$ is smaller than $-0.5$ because of the slow 
asymptotic nature of the curve near the critical value). 
The result of $g(j)$ thus obtained is shown as the
{\it long-dashed} line in \Fig{f1.eps} (b). 
The agreement between this model calculation and the simulation  
is good, verifying the self-consistency between the two PDFs in 
\Fig{f1.eps} and the relation between $(X,Y)$ given in 
Equation~(\ref{Eq:fit}).  
%This indicates that the scatter
%around the median relation is not the main cause of the bias.

We note that there are two constraints on $g(j)$,
\beq
\label{Eq:constraint}
\int_0^\infty g(j)dj=1, \quad\quad  \quad\quad \int_0^\infty jg(j)dj=1.
\eeq
\noindent
The empirical fit we adopted in Equation~(\ref{Eq:fit}) alone
does not satisfy the first constraint of
(\ref{Eq:constraint}); we must add the 
zero luminosity component, i.e., a Dirac delta function $\delta^D(j)$.
Hence the function $g(j)$ should be written as
\beq
\label{Eq:rewrite}
g(j) = A\delta^D(j) + g'(j), 
\eeq
and the first constraint reads
\beq
A + \int_0^{\infty} g'(j)dj=1,
\eeq
where $g'(j)$ corresponds to the light-emitting component discussed
above, i.e., the long-dashed line or the histogram itself in 
Figure~\ref{f1.eps} (b).
We find $A=0.6$ (see Section~\ref{Section:distribution}) and that 
the two constraints are satisfied for the model curve 
within an accuracy of $\sim 5 \%$ where 
we limit the integral to $\log j < 2.5$, the shot-noise limit of our
simulation. Needless to say, the simulation data automatically satisfy
the constraints by the definition of PDFs.

%%%%%%%%%%%%%%%%%%%%%%%%%%%%%%%%%%%%%%%%%%%%%%%%%%%%%%%%%%%%

\subsection{Distribution of Matter and Light}
\label{Section:distribution}

In this subsection, we study the distribution of mass and light from 
different angles and consider the relation to observations.
\Fig{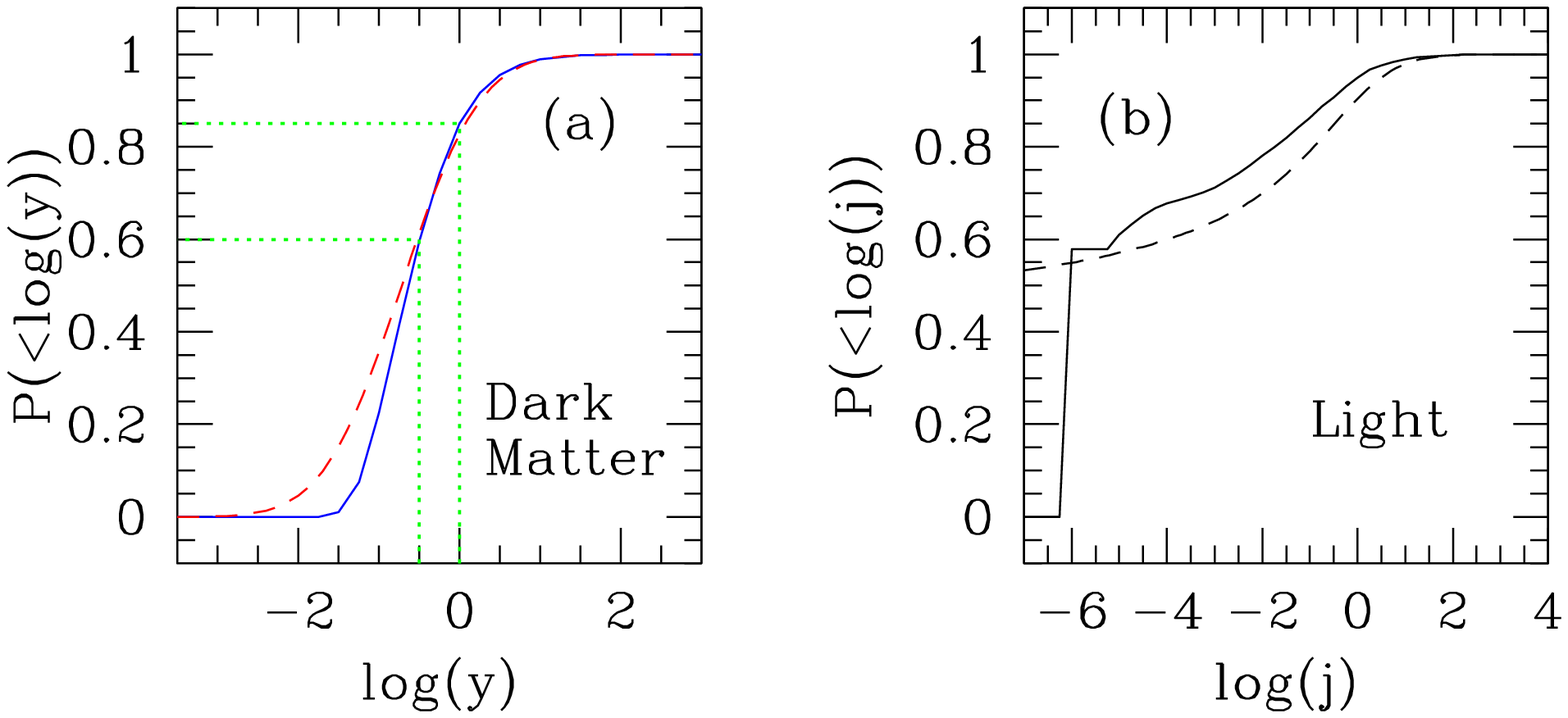}(a,b) shows the cumulative volume fraction of regions with mass 
density $<\log y$ and with light density $<\log j$, respectively; 
i.e., the probability [$P(<\log y, \log j)$] that a 
volume element has mass (light) density less than a given value of 
$\log y$ ($\log j$) in the abscissa. The {\it solid} line is computed 
from the simulated PDF, and the {\it short-dashed} line is obtained by 
integrating the model curve of \Fig{f1.eps}. We have seen that
no luminous objects are present in the region of $\log y<-0.5$.

\begin{figure}
\epsscale{1.0}
\plotone{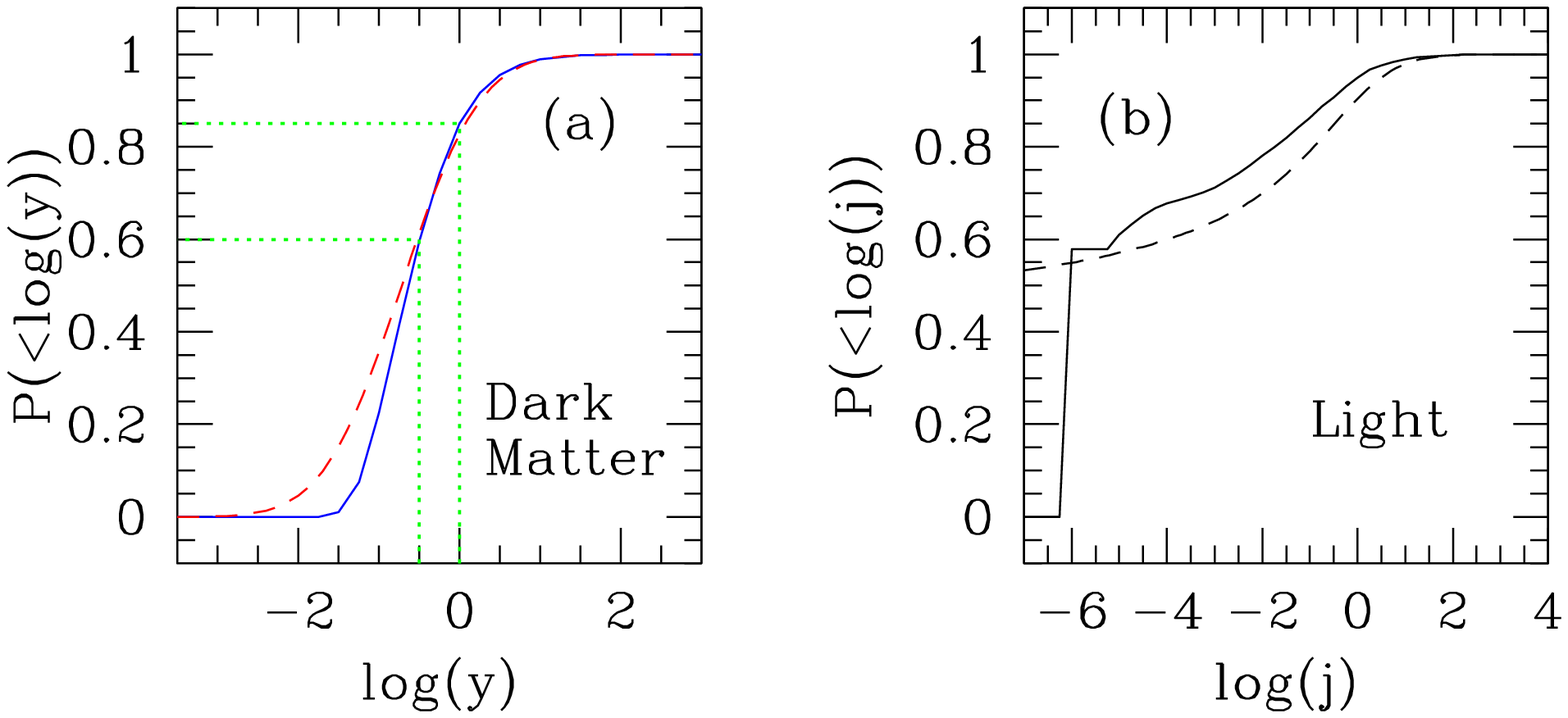}
\caption{Volume fraction of regions with mass density 
$<\log y$ (panel (a)) and with light density $<\log j$ (panel (b));
i.e, the probability that a volume element has mass (light) 
density less than a given value of $\log y$ ($\log j$) in the abscissa.
The {\it solid} line is computed from the simulated PDF, and the
{\it short-dashed} line is obtained by integrating the model fitted PDF
shown in \Fig{f1.eps}. The two dotted lines in panel (a) indicate that
the region  with $\log y<-0.5$ ($\log y<0$) occupies 60\% (85\%) of 
the volume.
\label{f3.eps}}
\end{figure}

Figure~\ref{f3.eps}(a) shows that few regions are emptier than $\log y<-2$:
i.e., the "voids", while they are completely dark, are not thoroughly 
devoided in mass \citep[cf.][]{Peebles01}. From this figure we see that 
the region  with $\log y<-0.5$ occupies 60\% of the volume (dotted line). 

We see in Figure~\ref{f3.eps}(b) that the light density exhibits 
a smooth distribution until it shows a jump at the faint end. The 
magnitude of the jump at $\log j=-6$ (solid line) means that spheres
in 60\% of volume contains no light; hence the value of $A$ in
Equation (7) is determined to be $A=0.6$.
The lower limit of $\log j$ is an artifact of the resolution of 
the simulation, but we consider that the realistic cutoff as determined 
by the Jeans and cooling conditions is rather close to this limit. 
The flattened distribution of $P(<\log j)$ towards the low luminosity 
end indicates that such a jump is present in reality. 

Let us consider voids defined in observations. Most surveys sample 
galaxies down to about 4 mag below $L^*$. By integrating a standard 
Schechter luminosity function (approximately $\phi^*=0.02h^3\mpc^{-3}$, 
$M^*=-20+5\log h$, and $\alpha=-1.25$) from $0.025 L^*$ 
(i.e. magnitude limit of $M=-16$) 
to infinity, we expect that a sphere of radius $R=1\mpc$ contains 
0.4 galaxies (here we assume $h=1$ for simplicity). This means that 
a sphere of luminosity $<0.01L^*$ cannot be observed in current standard
galaxy surveys.  Since the luminosity in a sphere of $R=1\mpc$ 
with the above Schechter function (with a magnitude limit of $0.025 L^*$) 
is $0.08L^*$, the observational threshold of light density parameter 
is $j_{\rm th}=(0.01/0.08)=0.125$, or 
$\log y_{\rm th}\approx 0$ using \Fig{f2.eps}. Namely, spheres in
the region of $\log y<0$ are observed as `dark'.  
Using \Fig{f3.eps}(a) this means that 85\% of volume is identified 
as voids (dotted line).

In order to see the mass in voids, we plot in \Fig{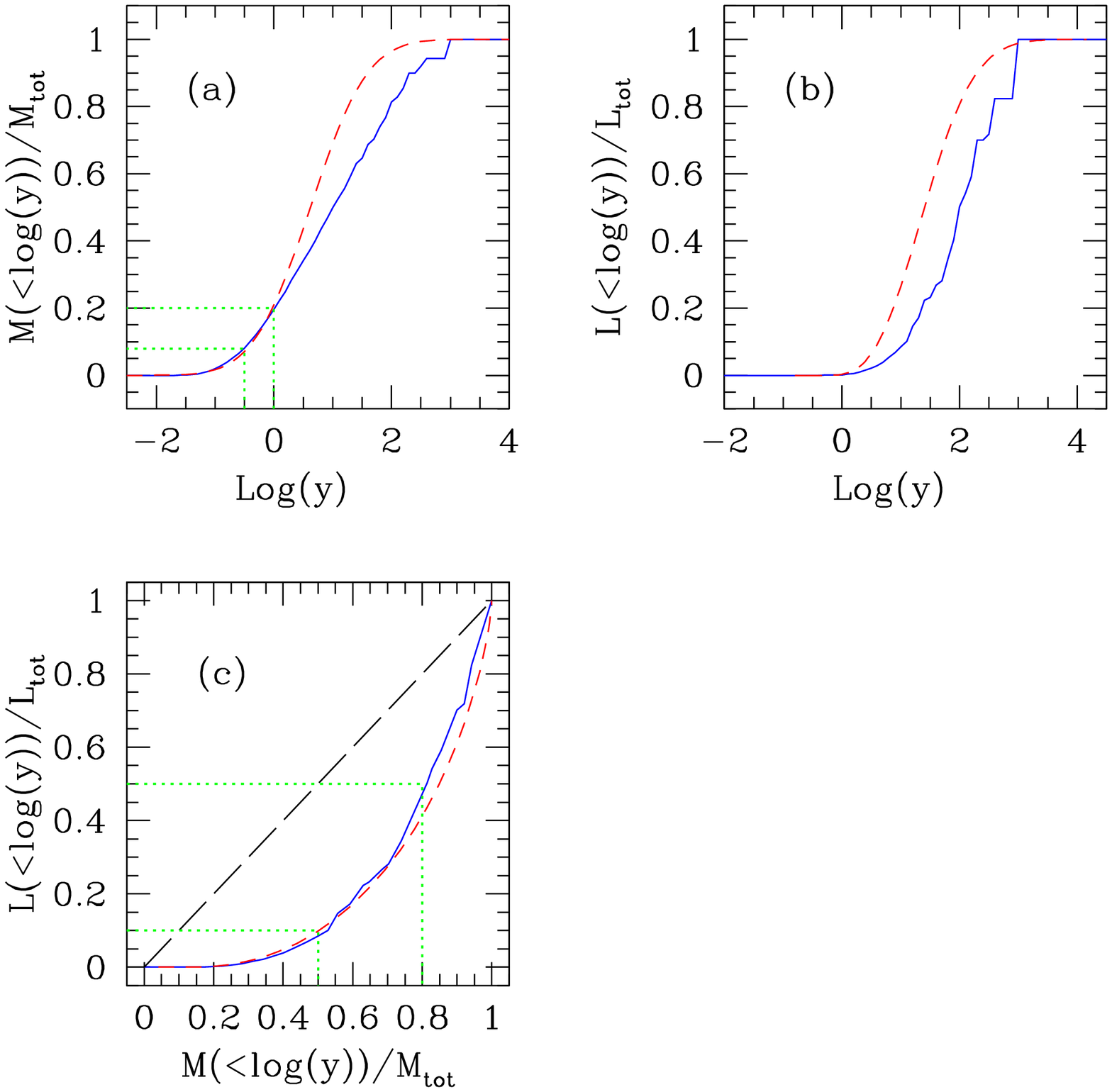} (a) 
the mass fraction and (b) light fraction  
that are contained in regions of $<\log y$, i.e.,
$\int_0^{y} y'f(y')dy'$ and $\int_0^{j} j'g(j')dj'$ for 
dark matter and light, respectively. 
Again, the {\it solid} line is computed from the simulated PDF, and the
{\it short-dashed} line is from the model curve.
The result is robust against the change in the lower limit of the 
integration. The discrepancy between the short-dashed line and the solid 
line is due to the difference in the shape of the simulated and the model
PDFs. 

\begin{figure}
\epsscale{1.0}
\plotone{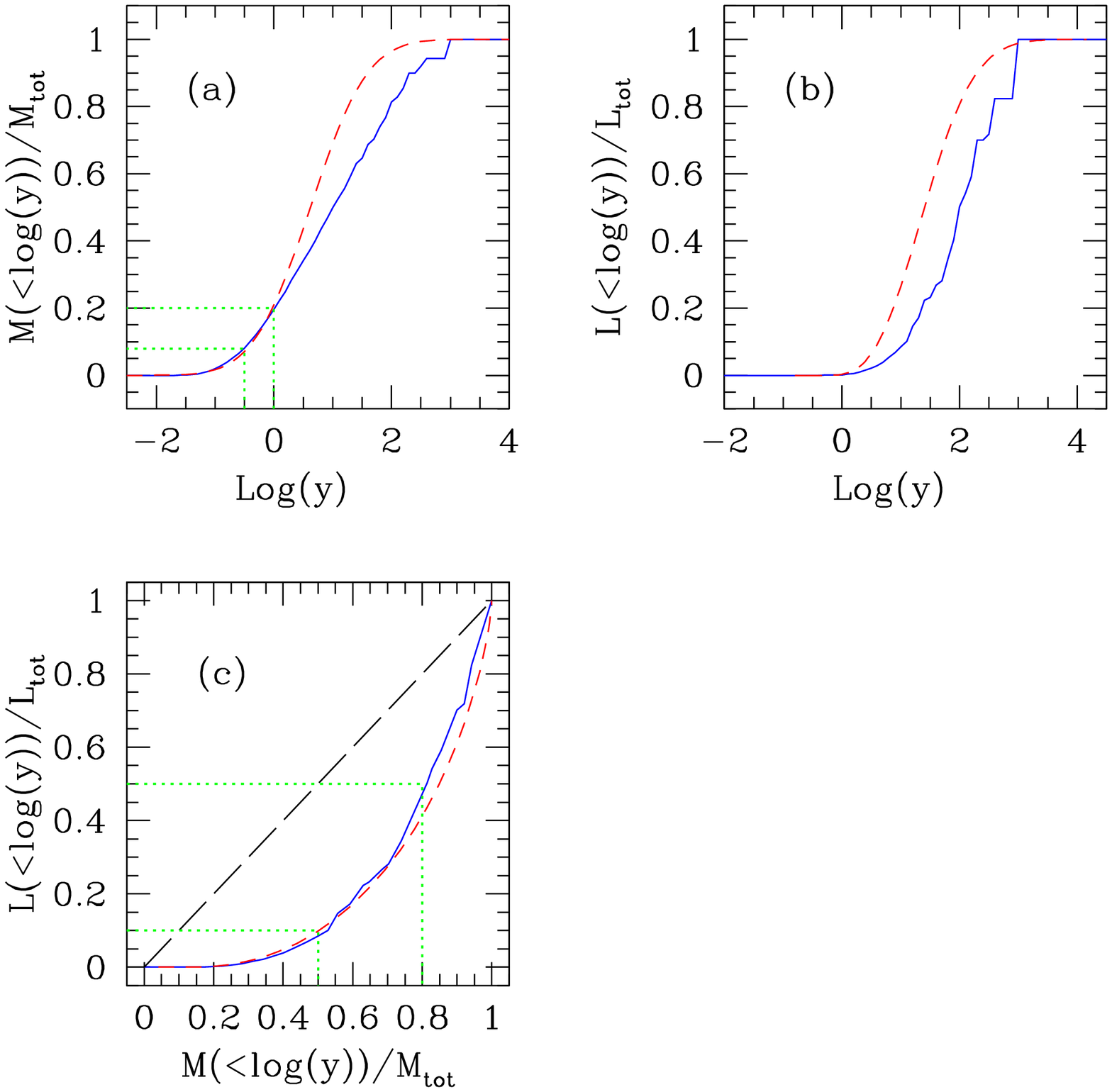}
\caption{Mass fraction (panel (a)) and light fraction 
(panel (b)) that is contained in regions of $<\log y$.
Note that the light is more concentrated in high density regions than the 
dark matter is. 
The {\it solid} line is computed from the simulated PDF, and the
{\it short-dashed} line is obtained by integrating the model
fitted PDFs: $\int_0^{y} y'f(y')dy'$ and $\int_0^{j} j'g(j')dj'$ for 
dark matter and light, respectively. The two dotted lines in panel (a)
indicate that the region with $\log y<-0.5$ ($\log y<0$) contain 10\% 
(20\%) of the total mass.
Panel (c) shows the mass and light 
fraction for the same values of $\log y$ as given in panels (a) \& (b).
The deviation from the diagonal {\it long-dashed} line indicates 
the difference in the manner that dark matter and light are distributed.
The two dotted lines in panel (c) indicate that 90\% of light comes from 
higher density regions where 50\% of mass is contained (or in other words, 
only 10\% of light is contained in lower density region where 50\% of mass 
is contained).
\label{f4.eps}}
\end{figure}

From \Fig{f4.eps}(a) we read that 8\% of mass (corresponding to
$\log y<-0.5$ where the sharp cutoff was observed in Figure~\ref{f1.eps}
) does not emit light at all, and observationally defined 
voids ($\log y<0$) contain 20\% of the total mass as indicated by 
the dotted lines.   
The relation between the mass fraction and the light fraction is
represented more clearly in Figure~\ref{f4.eps}(c), which indicates 
that 90\% of light comes from higher density regions where 50\% of mass 
is contained (or in other words, only 10\% of light is contained in lower 
density region where 50\% of mass is contained; see dotted lines).
The departure from the diagonal line in panel (c) shows the 
difference in the manner that dark matter and light are distributed 
in the universe.

%%%%%%%%%%%%%%%%%%%%%%%%%%%%%%%%%%%%%%%%%%%%%%%%%%%%%%%%%%%%%%%%%%%%%%

\section{Applications}
\label{Section:Applications}

\subsection{Mass-to-Light Ratio}
\label{Section:mlratio}

The result we have seen in the preceding section indicates that
the conventional evaluation of the the mass density 
from the luminosity density $\avg{\rho_L}$ and 
the mean mass-to-light ratio $\avg{M/L}$, 
as $\avg{\rho_m} = \avg{\rho_L} \avg{M/L}$ \citep{Peebles71},
underestimates the true value. 

\begin{figure}
\epsscale{1.0}
\plotone{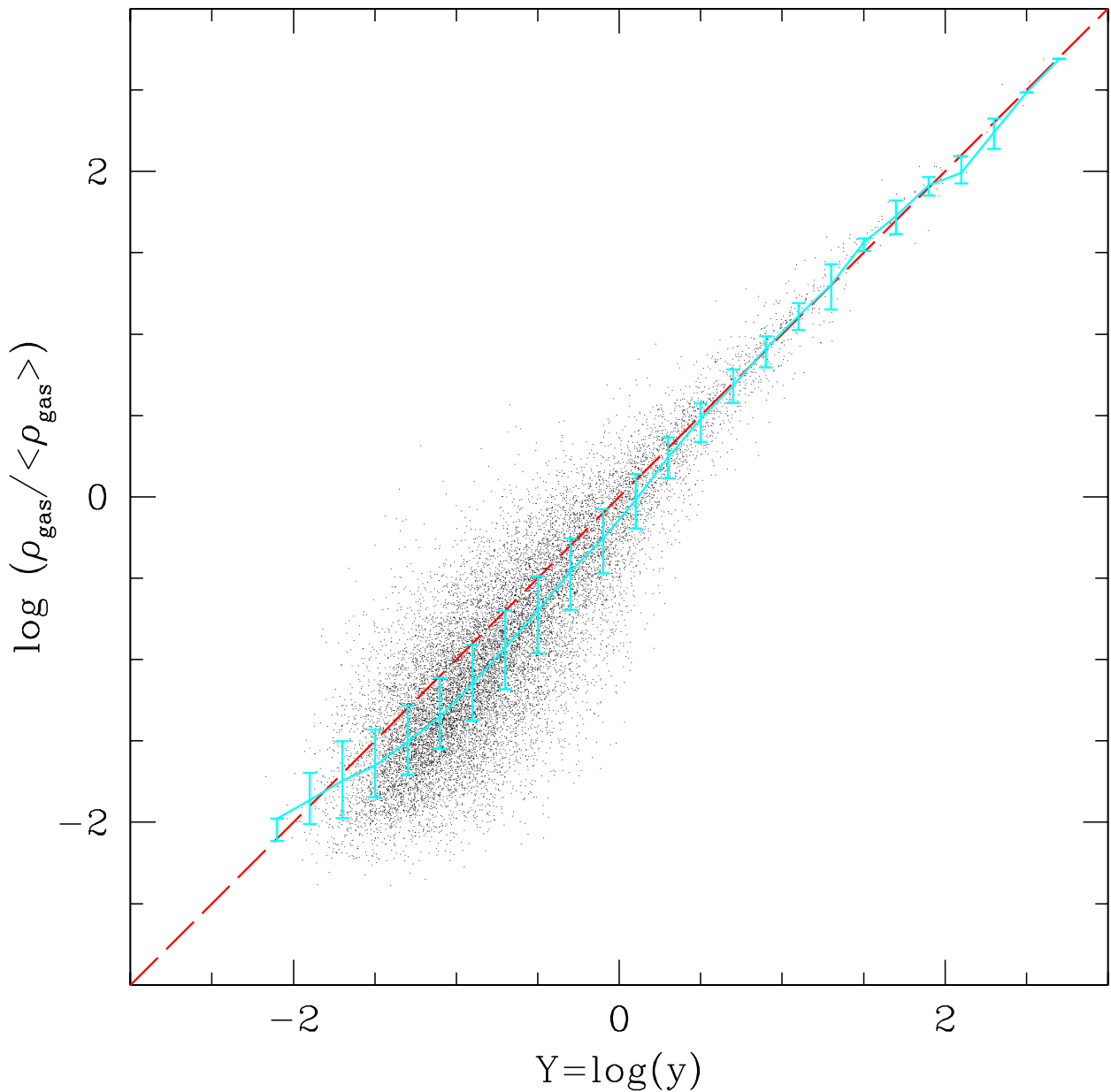}
\caption{Relation between gas overdensity $\rho_{gas}/\avg{\rho_{gas}}$
and total mass overdensity $Y=\log y$, where $y$ is the normalized 
matter density parameter. Each point in the figure represents a 
spherical volume element of radius $1\himpc$ in the simulation box. 
The {\it solid} line is the median values in each $Y$ bin, 
and the error bars are the quartiles on each side. The diagonal 
{\it long dashed} line shows the linear relation between the two quantities 
plotted.
\label{f5.eps}}
\end{figure}

If galaxy formation is considered as largely a local process, then 
it is expected to be primarily a function of local gas density and gas 
temperature. To a first approximation baryonic gas density should 
roughly trace the total mass density, and indeed this is verified in our
hydrodynamic simulation as shown in Figure~\ref{f5.eps}. Therefore, let us 
assume that galaxy formation is primarily a function of only the total 
mass density $\rho$. Namely, we take the normalized light density $j$ and
the local mass-to-light ratio $\mu (y) \equiv \frac{M}{L}(y) = 
\bar\rho y/j(y)$ as functions of $y$.

The volume averaged mean mass density can be calculated by
\beq
\avg{\rho}_V = \int_0^\infty \bar\rho y f(y)dy,
\eeq
and the volume averaged mean light density is 
\beq
\avg{j}_V = \int_0^\infty j(y) f(y)dy.
\eeq
So the appropriate global mean mass-to-light ratio would be
\beq
\label{Equation:mean_mu_vol}
\avg{\mu}_V \equiv \frac{\int_0^\infty \bar\rho y f(y)dy}{\int_0^\infty j(y)f(y)dy},
\eeq
which is defined so that, by construction,
\beq
\avg{\rho}_V =  \avg{\mu}_V \avg{j}_V.
\eeq

In observations one cannot observe the dark matter 
directly, and therefore, one samples only where the light is.
This could lead to an underestimate of mass since volume elements
in the universe are not treated equally. If we evaluate $\avg{\mu}_L$ 
observationally, $\avg{\mu}_{L, \rm obs}$, by studying only those
regions which emit light and weight them by the emitted light
$j$, then we are effectively evaluating the integral
\begin{eqnarray}
\label{Equation:mean_mu_lum}
\avg{\mu}_{L, \rm obs} &=& \frac{\int_0^\infty \bar\rho y j f(y)dy}{\int_0^\infty j^2 f(y)dy} \\
 &=& \frac{\int_0^\infty \bar\rho y j g(j)dj}{\int_0^\infty j^2 g(j)dj}.
\end{eqnarray}
To see how much underestimate could be caused by weighting by light, take the 
following simple model in which we assume that the mass-to-light ratio 
is constant ($=\mu_0$) in the galaxy forming regions. Namely we assume
that the light density is 
linearly proportional to the mass density, but that there are empty voids:
\begin{eqnarray}
j &=& 0 \qquad\quad \hbox{for $y<y_1$}\qquad \hbox{(``voids'')} \nonumber\\
j &=& \frac{\bar\rho}{\mu_0} y \qquad \hbox{for $y\ge y_1$}
\qquad \hbox{(``galaxy forming regions")}. 
\end{eqnarray}
This toy model is shown as {\it short-dashed} line in Figure~\ref{f2.eps}.
The mean mass-to-light ratio we defined in 
Eq.(\ref{Equation:mean_mu_vol}) becomes
\begin{eqnarray}
\avg{\mu}_V  &=& \mu_0 \frac{\int_0^\infty y f(y)dy}{\int_{y_1}^\infty yf(y)dy} \\
 &=& \mu_0 \left( 1+\frac{\int_0^{y_1} yf(y)dy}{\int_{y_1}^\infty yf(y)dy} \right).
\end{eqnarray}
The luminosity weighted average [Eq.(\ref{Equation:mean_mu_lum})] then
becomes simply
\beq
\avg{\mu}_L = \frac{\int_{y_1}^\infty \bar\rho y (\frac{\bar\rho}{\mu_0}y) f(y)dy}{\int_{y_1}^0 (\frac{\bar\rho}{\mu_0}y)^2 f(y)dy} 
= \mu_0.
\eeq
So the true volume weighted average $\avg{\mu}_V$ is larger than the 
luminosity weighted average $\avg{\mu}_{L, \rm obs}$ by a factor $\Gamma$:
\begin{eqnarray}
\label{Equation:Gamma}
\Gamma  &=& \frac{\avg{\mu}_V}{\avg{\mu}_L} = 1 + \frac{\int_{0}^{y_1} yf(y)dy}{\int_{y_1}^\infty yf(y)dy} > 1.
\end{eqnarray}
The ratio of the integrals in Eq.(\ref{Equation:Gamma}) is just 
the mass fraction in ``voids'' divided by that in ``galaxy forming 
regions''.  We estimated that the ``voids" contain 20\% of the total mass; 
this means that the correction factor will be $\Gamma=1.25$. 
This leads to an underestimate in $\Om$ by the same factor. 

The estimate of $\Om$ obtained by using $\avg{\rho_B}= (2.2\pm 0.4)\times 
10^8h \Lsunb\mpc^{-3}$ and $\avg{M/L}=290\pm35h$ \citep{McKay01} from 
gravitational lensing shears, $\Om=0.23\pm0.05$, should therefore be 
corrected to give $\Om=0.29\pm0.06$, in good agreement with the value
from the WMAP experiment: $\Om= 0.29\pm 0.07$ \citep{Spergel03}. 
The same correction also applies to similar estimates done to date,
which always show rather small values of $\Om$ compared to the estimate
of WMAP.
For example an estimate of \citet{Bahcall00}, $\Om=0.16\pm 0.05$,
using mass-to-light ratio as a function of scale is modified to
$0.20\pm 0.06$ after the correction. Similarly a low value of $\Om=0.15$
obtained by \citet[][equation (28)]{Fuku98} based on galaxy luminosities
should also receive an upward 25\% correction.

If the light PDF $g(j)$ is estimated directly from the observation,
one can numerically integrate the relation $dj/dy = f(y)/g(j)$ to 
obtain $j(y)$ using a cosmological simulation. With $j(y)$  
one can easily obtain the mass-to-light function $\mu(y)=\bar\rho y/j(y)$ 
and the correction factor  
\begin{eqnarray}
\Gamma = \frac{\avg{\mu}_V}{\avg{\mu}_L} = 
\left( \frac{\int_0^\infty yf(y)dy}{\int_0^\infty j(y) f(y)dy} \right) / \left( \frac{\int_0^\infty y j f(y)dy}{\int_0^\infty j^2 f(y)dy} \right). 
\end{eqnarray}

%%%%%%%%%%%%%%%%%%%%%%%%%%%%%%%%%%%%%%%%%%%%%%%%%%%%%%%%%%%%

\subsection{Cosmic Virial Theorem}

Let us now turn to the application to the ``cosmic virial theorem''.
\citet{Peebles80, Peebles93} shows that, if one assumes that 
galaxies trace mass, the ``cosmic virial theorem'' which relates the 
relative velocity dispersion and the mean mass density can be 
approximated by 
\beq
\Omega_m \sim \frac{2(3-\gamma)}{3}\frac{\sigma(r)^2}{H_0^2r_0^{\gamma}r^{2-\gamma}}.
\label{Eq:Om}
\eeq
To derive this equation, $\xi_{gg}(r) = \xi_{gm}(r)$ is assumed, where 
$\xi_{gg}(r)=(r/r_{0,gg})^{-\gamma}$ is the galaxy-galaxy 
correlation function, and $\xi_{gm}(r)=(r/r_{0,gm})^{-\beta}$ is 
the galaxy-mass cross-correlation function.

As we showed earlier, galaxies do not trace mass well.  Therefore 
if instead one were to use the galaxy-mass cross-correlation function 
in deriving the above relation, the value of $\Om$ would have been 
larger or smaller by a factor of 
\beq
\Gamma' \equiv \frac{\Omega_{m,gm}}{\Omega_{m,gg}}=\left (\frac{3-\beta}{3-\gamma}\right ) \left (\frac{r_{0,gg}^{\gamma}}{r_{0,gm}^{\beta}} \right ),
\label{Eq:correct}
\eeq
where $r$ is measured in units of $\himpc$ and the right-hand-side of 
Eq.~(\ref{Eq:Om}) is measured at $r=1\himpc$.

The values of $\gamma$ and $r_{0,gg}$ are observationally known ($\gamma\simeq 1.8$, $r_{0,gg}\simeq 5\himpc$) and 
$\beta$ and $r_{0,gm}$ can be estimated in the linear regime as follows:
\begin{eqnarray}
\xi_{gm} &=& \avg{\rp{1+\rp{\frac{\drho}{\rho}}_g} \rp{1+\rp{\frac{\drho}{\rho}}_m}} - 1 \\
&=& \avg{\rp{\frac{\drho}{\rho}}_g \rp{\frac{\drho}{\rho}}_m } \\
&=& \frac{1}{\kappa} \avg{\rp{\frac{\drho}{\rho}}_g \rp{\frac{\drho}{\rho}}_g } \\
&=& \frac{1}{\kappa}~\xi_{gg} = \frac{1}{\kappa}\rp{\frac{r}{r_{0,gg}}}^{-\gamma}, \label{Eq:compare}
\end{eqnarray}
where we have assumed a relation 
\beq
\frac{\rho_g}{\avg{\rho_g}} = \rp{\frac{\rho_m}{\avg{\rho_m}}}^{\kappa},
\eeq
and therefore, 
\beq
\rp{\frac{\drho}{\rho}}_g = \kappa \rp{\frac{\drho}{\rho}}_m.
\eeq
The analytic fit that we adopted in \S~\ref{Section:XYrelation} 
Eq.~(\ref{Eq:fit}) implies $\kappa\simeq a\simeq 1.4$ for the 
domain of $0\leq \log y\leq 2$
that concerns us.
By comparing Eq.~(\ref{Eq:compare}) and the original functional form for 
$\xi_{gm}$, one obtains
\beq
r_{0,gm} = \kappa^{-\frac{1}{\gamma}} ~ r_{0,gg}.
\eeq
For any value of $\kappa$ larger than unity will therefore gives
\beq
\beta = \gamma \quad {\rm and}\quad  r_{0,gm} < r_{0,gg}.
\eeq

Using these relations in Eq.~(\ref{Eq:correct}), the 
matter density of the universe estimated by Peebles by this method 
($\Omega_m\sim 0.2$) should be increased by a factor of $\Gamma'=1.2$, which 
yields a corrected value of $\Omega_m=0.24$.
The agreement with the WMAP result is improved by this correction.

%%%%%%%%%%%%%%%%%%%%%%%%%%%%%%%%%%%%%%%%%%%%%%%%%%%%%%%%%%%%%%%%%%%%%%

\section{Discussion and  Conclusions} 

Very little work has been done on the optical probability distribution
function (PDF): the probability of finding a region (of some given size)
containing a specified range of optical light output from galaxies 
within that region. Even less has been done in relating the optical
PDF with the matter (or dark matter) PDF. 

What is known is that the low density regions (optical ``voids'')
are relatively unpopulated by galaxies \citep{Kirshner81, Rood88},
and consequently are optically faint as compared from the dark matter
simulations of the currently popular concordance model. 
If this is correct, then ``bias'' --- the ratio of galaxies to dark 
matter --- is different in high and low density regions. And more 
seriously, our observational tracers of cosmic structure are giving
us a view of the universe that is inherently contaminated by selection
bias.

We compute the effects in a TVD cosmic hydrodynamic simulation
and then estimate the error that would have been involved, had these
simulations been observed and analyzed using conventional techniques.
Focusing on the global value of $\Om$ (which of course is known for 
the simulated universe), we can estimate the corrections, and when
applying these corrections we obtain increased values of $\Om$
to the range $0.2-0.3$, which is in better accord with the CMB 
determinations.

\begin{figure}
\epsscale{1.0}
\plotone{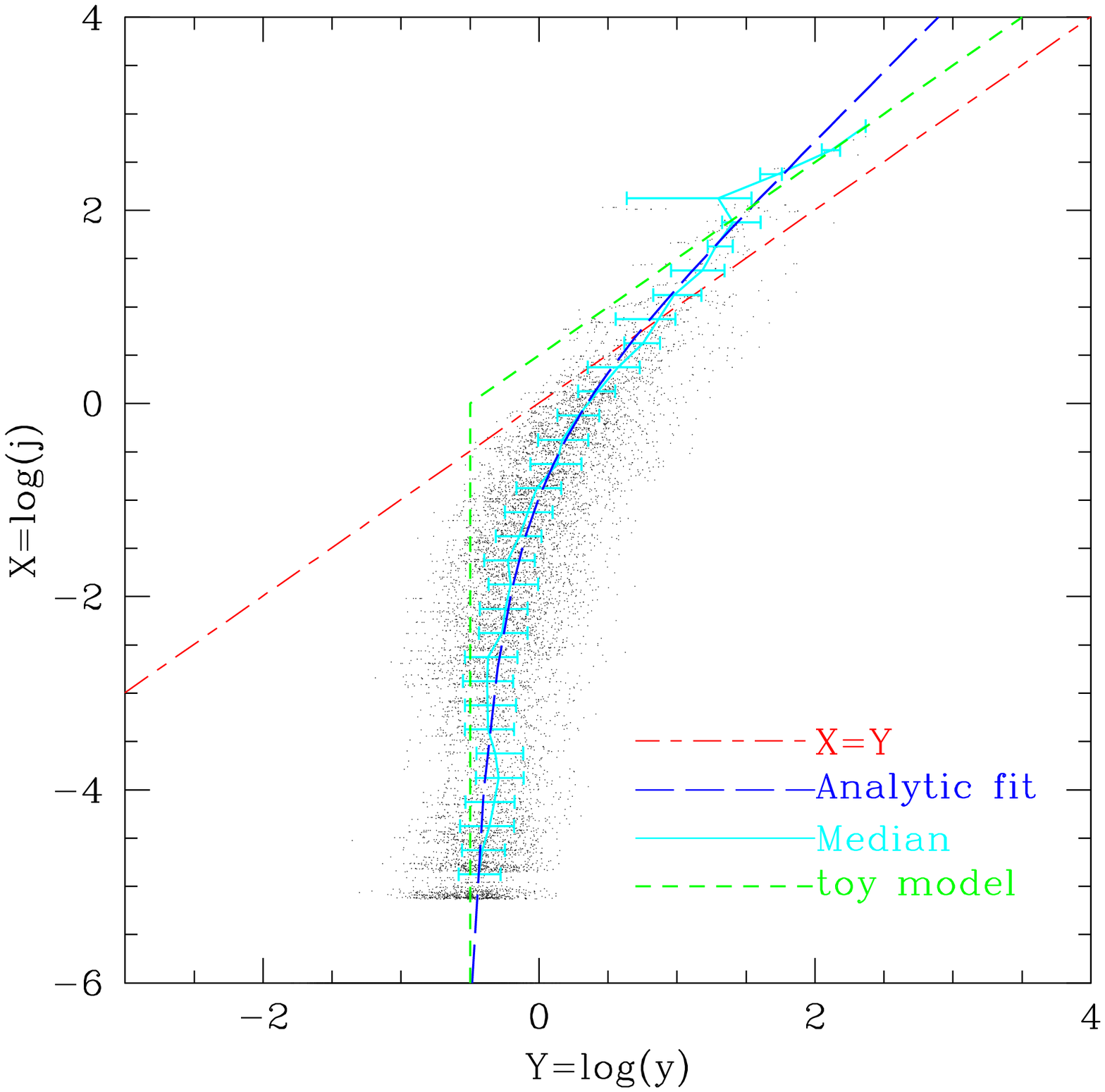}
\caption{Relation between $X=\log j$ and $Y=\log y$, where $j$ and $y$ 
are the normalized light and matter density parameter, respectively.
Each point in the figure represents a tophat sphere of radius $R=1\himpc$ 
placed in the simulation box arbitrarily. 
Notice that in the significant fraction of the volume having mass density
less than $1/3$ of the average value, there is effectively no light 
density, i.e., $M/L \rightarrow \infty$. 
{\it Solid} line: median values in each $X$ bin, and the error bars are the 
quartiles on each side. {\it Long-short dashed} line: linear relation 
$X=Y$.  {\it Long dashed} line: empirical fit explained in 
\S~\ref{Section:XYrelation}.  {\it Short-dashed} line: toy model 
described in \S~\ref{Section:mlratio} with a cutoff at $\log y = -0.5$. 
\label{f2.eps}}
\end{figure}

The relation between mass PDF and light PDF also gives us insight
concerning the nature of the void phenomenon, and enables us
to estimate the mass and volume fractions occupied by voids. Our
result indicates that no galaxies can be present in the middle of 
voids: sub-luminous galaxies are distributed in the edge of voids,
whereas luminous galaxies reside only in high-density filaments.   
The quantitative basis for these statements will be presented in 
a subsequent paper, but the concept is as follows. A large void 
can be considered as a small piece of a lower density universe.
The implies a shift to a lower characteristic halo mass in such 
regions and a shift in the horizontal scale in Figures~(\ref{f2.eps}) 
and (3a) to domains in which galaxy formation is less efficient.  
At the centers of such voids all galaxy formation is suppressed 
and at the edges high luminosity galaxies (from higher mass halos) 
are suppressed more than lower luminosity systems. A recent numerical
study by \citet{Got03} supports this view.

Future work using large-scale surveys can provide the empirical data
needed to compute the optical PDF properly. When this is combined
with detailed numerical simulations of the dark matter, we will be 
able to self-consistently evaluate these effects better and use the 
so determined relation between light density and mass density in our
analysis of observational data.

%%%%%%%%%%%%%%%%%%%%%%%%%%%%%%%%%%%%%%%%%%%%%%%%%%%%%%%%%%%%%%%%%%%%%%

\acknowledgments
We thank Michael Strauss for useful comments on the manuscript.
K.N. thanks for the hospitality of the Institute of Astronomy,  
University of Cambridge where a part of this work was completed. 
This work was supported in part by grants AST~98-03137 and 
ASC~97-40300 in Princeton, and Grant in Aid of the Ministry of 
Education 13640265 in Japan.

%%%%%%%%%%%%%%%%%%%%%%%%%%%%%%%%%%%%%%%%%%%%%%%%%%%%%%%%%%%%%%%%%%%%%%

%%%%%%%%%%%%%%%%%%%%%%%%%%%%%%%%%%%%%%%%%%%%%%%%%%%%%%%%%%%%%%%%%%%%%%

%%%%%%%%%%%%%%%%%%%%%%%%%%%%%%%%%%%%%%%%%%%%%%%%%%%%%%%%%%%%%%%%%%%%%%


\begin{thebibliography}{}
\bibitem[Bahcall et al.(2000)]{Bahcall00}
	Bahcall, N. A., Cen, R., Dav\'{e}, R., Ostriker, J. P., \& Yu, Q.
	2000, ApJ, 541, 1

\bibitem[Benson, et al.(2003)]{Benson03} Benson, A., Hoyle, F., 
	Torres, F., Vogeley, M. S., 2003, MNRAS, 340, 160

\bibitem[Blanton \etal(1999)]{Blanton99} Blanton, M., Cen, R., 
	Ostriker, J. P., Strauss, M. A., 1999, \apj, 522, 590 

\bibitem[Bruzual \& Charlot(1993)]{BC93} Bruzual, A. G. \& Charlot, S., 
        1993, \apj, 405, 538

\bibitem[Cen \& Ostriker(1993)]{CO93} Cen, R., \& Ostriker, J.P. 
	1993, \apj, 417, 404

\bibitem[Cen et al.(2002)]{CO02} Cen, R., Ostriker, J. P., Prochaska, J. X., \& Wolfe, A. M., 2002, preprint (astro-ph/0203524)

\bibitem[Coles \& Jones(1991)]{Coles91} Coles, P. \& Jones, B. 1991, 
	\mnras, 248, 1

\bibitem[Dekel \& Lahav(1999)]{Dekel99} Dekel, A. \& Lahav, O. 1999, 
	\apj, 520, 24

\bibitem[Efstathiou et al.(1990)]{Efstathiou90} Efstathiou, G. et al. 
	1990, \mnras, 247, 10

\bibitem[Fukugita, Hogan, \& Peebles(1998)]{Fuku98} Fukugita, M., 
	Hogan, C. J., \& Peebles, P. J. E. 1998, ApJ, 503, 518

%\bibitem[Gaztanaga(1994)]{Gaz94} Gaztanaga, E., 1994, MNRAS, 268, 913

\bibitem[Gottl\"{o}eber et al.(2003)]{Got03} Gottl\"{o}eber, S., Lokas, E., Klypin, A., 
	Hoffman, Y. 2003, MNRAS, submitted (astro-ph/0305393)

%\bibitem[Groth \& Peebles(1977)]{Groth77} Groth, E. J. \& 
%	Peebles, P. J. E. 1977, \apj, 217, 385

\bibitem[Kayo, Taruya, \& Suto(2001)Kayo et al.]{Kayo01} Kayo, I., 
	Taruya, A., \& Suto, Y. 2001, \apj, 561, 22

\bibitem[Kirshner et al.(1981)]{Kirshner81} Kirshner, R. P., Oemler, A., Jr.,
	Schechter, P. L., \& Shectman, S. A., 1981, ApJ, 248, L57

\bibitem[Kofman et al.(1994)]{Kofman94} Kofman, L., Bertschinger, E., 
	Gelb, J. M., Nusser, A., \& Dekel, A. 1994, \apj, 420, 44

\bibitem[McKay et al.(2001)]{McKay01} McKay, T. A. et al., preprint
        astro-ph/0108013

%\bibitem[Mo, Jing, \& White(1997)]{Mo97} Mo, H. J., Jing, Y. P., \& 
%	White, S. D. M., 1997, MNRAS, 284, 189

\bibitem[Nagamine(2002)]{Nagamine02} Nagamine, K. 2002, \apj, 564, 73

\bibitem[Nagamine, Fukugita, Cen, \& Ostriker(2001b)Nagamine et al.]{Nagamine01b}
	Nagamine, K., Fukugita, M., Cen, R., \& Ostriker, J. P. 2001, 
	\mnras, 327, L10

\bibitem[Nagamine, Fukugita, Cen, \& Ostriker(2001a)Nagamine et al.]{Nagamine01a}
	Nagamine, K., Fukugita, M., Cen, R., \& Ostriker, J. P. 2001, 
	\apj, 558, 497

\bibitem[Peebles(1971)]{Peebles71} Peebles, P. J. E. 1971, 
	{\it Physical Cosmology}, Princeton University 
	Press, Princeton, New Jersey

\bibitem[Peebles(1980)]{Peebles80} Peebles, P. J. E. 1980, 
	{\it The Large-Scale Structure of the Universe}, Princeton 
	University Press, Princeton, New Jersey, \S~74

\bibitem[Peebles(1993)]{Peebles93} Peebles, P. J. E. 1993, 
	{\it Principles of Physical Cosmology}, Princeton University 
	Press, Princeton, New Jersey, p.481

\bibitem[Peebles(2001)]{Peebles01} Peebles, P. J. E. 2001, ApJ, 557, 495

\bibitem[Rood(1988)]{Rood88} Rood, H. J., 1988, ARA\&A, 26, 245

\bibitem[Ryu et al.(1993)]{Ryu} Ryu, D., Ostriker, J. P., Kang, H., Cen, R.
	1993, ApJ, 414, 1

\bibitem[Sigad, Branchini, \& Dekel(2000)]{Sigad00} Sigad, Y., Branchini, E.,
	Dekel, A. 2000, ApJ, 540, 62

\bibitem[Spergel et al.(2003)]{Spergel03} Spergel, et al. 2003, ApJ, accepted
	(astro-ph/0302209)

%\bibitem[Szapudi, et al.(2002)]{Szapudi02} Szapudi, et al., 2002, ApJ, 570, 75

\bibitem[Taylor \& Watts(2000)]{Taylor00} Taylor, A. N. \& Watts, P. I. R.
	2000, \mnras, 314, 92 

%\bibitem[Totsuji \& Kihara(1969)]{Totsuji69} Totsuji, H. \& Kihara, T.
%	1969, PASJ, 21, 221

\bibitem[Vogeley, Geller, \& Huchra(1991)]{Vogeley91} Vogeley, M. S., 
	Geller, M. J., \& Huchra, J. P., 1991, ApJ, 382, 44
 
\bibitem[Vogeley, et al.(1994)]{Vogeley94} Vogeley, M. S., 
	Geller, M. J., Changbom, P., \& Huchra, J. P., 1994, ApJ, 108, 745

\end{thebibliography}
\end{document}